\documentclass[aps,prb,twocolumn,showpacs]{revtex4}
\usepackage{graphicx,amssymb,amsfonts,amsmath}

\begin{document}

\title{Phonon-mediated tuning of instabilities in the Hubbard model at 
half-filling}
\author{F. D. Klironomos}
    \email{fkliron@physics.ucr.edu}
    \affiliation{Physics Department, University of California, Riverside,
CA 92521}
\author{S.-W. Tsai}
    \email{swtsai@physics.ucr.edu}
    \affiliation{Physics Department, University of California, Riverside, 
CA 92521}

\date{\today}


%%%%%%%%%%%%%%%%%%%%%%%%%%%%%%%%%%%%%%%%%%%%%%%%%%%%%%%%%%%%%%%%%%%%%%%%%%%%%%%
%  Convenient definitions  
%%%%%%%%%%%%%%%%%%%%%%%%%%%%%%%%%%%%%%%%%%%%%%%%%%%%%%%%%%%%%%%%%%%%%%%%%%%%%%%
\newcommand{\vc}[1]{ {\mathbf #1} }
\newcommand{\psid}{\psi^{\dagger}}
\newcommand{\phid}{\phi^{\dagger}}
\newcommand{\intkw}{\int_{\vc{k}\omega}\!}
\newcommand{\intkiwi}{\int_{\vc{k}_i\omega_i}\!}
\newcommand{\intqW}{\int_{\vc{q}\Omega}\!}
\newcommand{\intk}{\int_{\vc{k}}\!}
\newcommand{\ksik}{\xi_{\vc{k}}}
\newcommand{\uk}{u(k_1k_2k_3)}
\newcommand{\Ll}{\Lambda_\ell}
\newcommand{\lp}{{\ell_{p}}}
\newcommand{\lpp}{{\ell_{q_{pp}-p}}}
\newcommand{\lppt}{{\tilde{\ell}_{pp}}}
\newcommand{\lph}{{\ell_{p+q_{ph}}}}
\newcommand{\lpht}{{\tilde{\ell}_{ph}}}
\newcommand{\lphp}{{\ell_{p+q'_{ph}}}}
\newcommand{\lphpt}{{\tilde{\ell}'_{ph}}}
%%%%%%%%%%%%%%%%%%%%%%%%%%%%%%%%%%%%%%%%%%%%%%%%%%%%%%%%%%%%%%%%%%%%%%%%%%%%%%%

\begin{abstract}

We obtain the phase diagram of the half-filled two-dimensional Hubbard model
on a square lattice in the presence of Einstein phonons. We find that the 
interplay between the instantaneous electron-electron repulsion and 
electron-phonon interaction leads to new phases. In particular, a 
d$_{x^2-y^2}$-wave superconducting phase emerges when both anisotropic 
phonons and repulsive Hubbard interaction are present. For large 
electron-phonon couplings, charge-density-wave and s-wave superconducting
regions also appear in the phase diagram, and the widths of these regions 
are strongly dependent on the phonon frequency, indicating that retardation 
effects play an important role. Since at half-filling the Fermi surface is 
nested, spin-density-wave is recovered when the repulsive interaction 
dominates. We employ a functional multiscale renormalization-group 
method\cite{SWTsai2005} that includes both electron-electron and 
electron-phonon interactions, and take retardation effects fully into account.

\end{abstract}

\maketitle

\section{Introduction}

The renormalization-group (RG) approach to interacting 
fermions\cite{Shankar1994} has recently been extended to include the study of
interacting fermions coupled to bosonic modes, such as phonons\cite{SWTsai2005}. 
Experimental evidence indicates that in many strongly correlated systems, such 
as organic conductors and superconductors\cite{organics}, 
cuprates\cite{cuprates1,cuprates2}, cobaltates\cite{cobaltates}, and conducting 
polymers\cite{polymers}, both electron-electron interactions and phonons may 
play an important role. Also, recent advances in the field of cold atoms have 
made possible the creation of fermion-boson mixtures on artificial lattices. 
In these mixtures the fermions interact through instantaneous on-site 
repulsion, and when the bosonic atoms condense, there is an additional retarded 
attractive interaction mediated by the fluctuations of the bosonic condensate; 
a scenario similar to that of electron and phonon interaction in solid state 
systems\cite{Matley2006}. The physics of the interplay between the repulsive 
electron-electron (e-e) and attractive electron-phonon (e-ph) interaction is not 
well understood and fundamental questions arise that include a full understanding 
of retardation effects and whether competition/cooperation between these 
interactions can lead to new phases.

\begin{figure}[t]
\includegraphics[angle=-90,totalheight=5.5cm,width=7.0cm,viewport=5 5 570
710,clip]{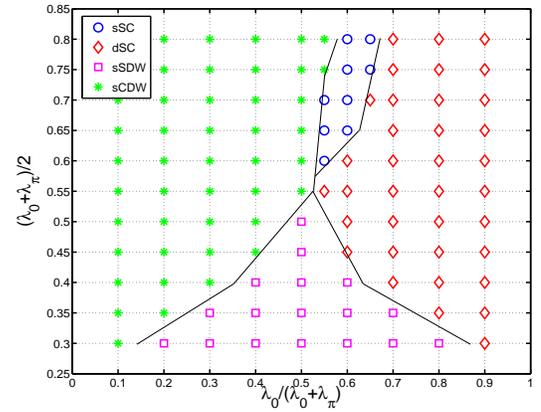} 
\caption{(Color online) Phase diagram for Einstein phonons of frequency 
$\omega_E=1.0$. Four phases involving antiferromagnetism (sSDW) (purple 
squares), charge density wave (sCDW) (green stars) and s-wave (sSC) (blue 
circles) and d-wave (dSC) (red rhombs) superconductivity compete
in the vicinity where the average phononic strength $\bar{\lambda}$ approaches
the bare on-site repulsion $u_0=0.5$. The lines distinguishing the different 
domains are guides to the eye.}
\label{Fig:PhaseDiagram:1.0}
\end{figure}
One of the most extensively studied model is the two-dimensional repulsive 
Hubbard model\cite{Schulz1987,Hubbard-RG,Gonzalez1996,Senechal2005}, which in 
the absence of phonons and at half-filling becomes an antiferromagnet 
due to nesting in the Fermi surface (FS) which drives an s-wave spin density wave 
(sSDW) instability. In the presence of isotropic phonons and in the strongly 
coupled regime, it has been shown that the s-wave charge density wave 
instability dominates over antiferromagnetism and coexists with s-wave
superconductivity (sSC)\cite{Berger1995}. When anisotropic phonons are present 
and full retardation is taken into account the phase diagram in parameter space 
associated with this model is summarized in Fig.~(\ref{Fig:PhaseDiagram:1.0}),
which reproduces the usual sSDW and sCDW instabilities but additionally a 
d$_{x^2-y^2}$-wave superconducting phase (dSC). Therefore, in this work we 
show that this dSC phase which appears at half-filling, is a result of 
{\it cooperation} between the two interactions. We stress that the role of the 
phonons goes beyond simply creating an effective attractive force between the 
electrons. Retardation plays a crucial role in generating the new phases, and 
the size of each region depends on the phonon frequency. This model can also 
be used to study various other systems where the e-ph interaction is present such 
as the quasi-2D $\kappa$-(BEDT-TTF)$_2$-X materials, which are stoichiometric 
with a fixed density of one electron per BEDT-TTF dimer\cite{kappa}, or as we 
mentioned in the beginning, for fermion-boson optical lattice mixtures (where 
there is experimental control on the number of fermions per lattice point) deep 
in the bosonic condensate phase. This study is by no means complete in its 
results since it is limited to the weak coupling regime but is intended to 
chart the vast phase space of these type of systems with some preliminary 
results that will help arrange the important underlying physics and contribute 
to a better understanding of the competition and cooperation between the 
interactions involved.

This paper is organized as follows. In section II we introduce the theoretical 
model and the RG method of analysis we employ. In section III we provide our 
results and draw the phase diagrams associated with the different orders in the 
system. In section IV we discuss the summary and the basic physics our work has 
highlighted and finally in the appendix we provide more details for the RG flow 
of the couplings and susceptibilities for the reader that is interested on the 
theoretical details.

\section{Theoretical Model}

The approach we employ in this study is based on a general RG analysis of a system
of electrons recently expanded to involve the coupling of the electrons with phonons 
as well\cite{SWTsai2005}. We use a generic model of electrons on a Hubbard lattice 
at half-filling interacting through the repulsive Coulomb interaction and being
isotropically and anisotropically coupled to dispersionless bosonic excitations 
(Einstein phonons). The Hamiltonian associated with this type of system is 
\begin{eqnarray}
H \!\!\!&=&\!\!\! \sum_{\vc{k},\sigma} \ksik
  c_{\vc{k},\sigma}^{\dagger} c_{\vc{k},\sigma} + \!\!\! \sum_{\vc{k}_1,
  \vc{k}_2, \vc{k}_3} \!\! \sum_{\sigma} u_0 c_{\vc{k}_3,\sigma}^{\dagger}
  c_{\vc{k}_4 ,-\sigma}^{\dagger} c_{\vc{k}_2,-\sigma} c_{\vc{k}_1,\sigma}
  \nonumber\\ 
&&+ \omega_E\sum_{\vc{q}} b_{\vc{q}}^{\dagger} b_{\vc{q}} + \sum_{\vc{q},\vc{k}}
  g(q) c_{\vc{k}+\vc{q},\sigma}^{\dagger} c_{\vc{k},\sigma} (b_{\vc{q}} +
  b_{-\vc{q}})
\label{H_model}
\end{eqnarray}
where $c_{\vc{k},\sigma}^{\dagger}$($c_{\vc{k},\sigma}$) is the creation
(annihilation) operator of an electron with momentum $\vc{k}$ and spin
$\sigma$, and $b_{\vc{q}}^{\dagger}$ ($b_{\vc{q}}$) is the corresponding
creation (annihilation) operator of a phonon with momentum $\vc{q}$. Also,
$\ksik = - 2 t (\cos k_x + \cos k_y)$ is the non-interacting
electron energy at half-filling, $\omega_E$ is the Einstein frequency, 
$u_0$ is the e-e on-site repulsion, while the e-ph coupling $g(q)$ is taken to 
be momentum dependent. Momentum conservation (up to reciprocal lattice vectors)
implies $\vc{k}_4=\vc{k}_1+\vc{k}_2-\vc{k}_3$. Going to the path-integral 
formulation, the bosonic fields can be integrated out exactly\cite{SWTsai2005}
to give an effective e-e interaction
\begin{equation}
U(k_1,k_2,k_3)=u_0-\lambda_{|\vc{k}_1-\vc{k}_3|}\frac{\omega_E^2}{\omega_E^2
+(\omega_1 -\omega_3)^2},
\label{U_eff}
\end{equation}
where $k_i\equiv(i\omega_i,\vc{k}_i)$, and $\lambda_q \equiv 2 g(q)^2/\omega_E$.
The bare interaction therefore contains an instantaneous on-site Coulomb
repulsion term and a retarded attractive part. This is the interaction used in 
the functional RG analysis in this work.

\subsection{RG flow of the couplings}

The general idea of RG theory is to integrate out self-consistently the 
degrees of freedom in the electronic system  in successive steps (called RG 
time), so that the electronic interaction becomes renormalized with each 
RG time step until it diverges, which is a signature of an instability. The 
approximate character of this method lies in the self-consistent evaluation of 
the renormalization of the electronic interactions, where bare values are
assumed to be smaller in strength than the band-width of the available 
electronic energies. The same approximation applies for the electron-phonon 
coupling strength as well (weak coupling) which results for the parameters of 
Eq.~(\ref{H_model}) to obey $u_0,\omega_E,g(q)\ll \Lambda_0=4t$. 
We also remain within one-loop accuracy\cite{Shankar1994} which 
is adequate to capture the essential physics, but we expand the analysis to 
include dynamics (frequency dependence) in the electronic couplings. 
In previous work\cite{Zanchi2000,Halboth2000,Binz2003} the couplings involved in
the RG study were frequency-independent which led to self-energy corrections 
at one-loop that were neglected in order to keep the total number of electrons 
fixed. In this work, the implicit frequency dependence generates an imaginary part
in the self-energy $\Sigma_{\ell}(k)$ which we calculate in order to use 
the full cut-off dependent electron propagator, given by
\begin{equation}
C_\ell(p)=\frac{\theta(|\xi_\vc{p}|-\Lambda_\ell)}{i\omega-\xi_\vc{p}
-\Sigma_\lp(p)},
\label{C_ell}
\end{equation}
where $\theta(x)$ is the Heaviside step function, $\Lambda_\ell$ is the value 
of the RG cutoff at the RG time $\ell=\ln(\Lambda_0/\Lambda_{\ell})$ and 
$\ell_p=\ln(\Lambda_0/|\xi_\vc{p}|)$, where $\Lambda_0$ is the initial cut-off 
corresponding to half the bandwidth ($4t$) as we defined above. 
As we have already mentioned the RG flow equations are evaluated at 
one-loop\cite{Shankar1994} accuracy using the general Polchinski 
equation\cite{Polchinski1984} applied for this specific lattice 
model\cite{Zanchi2000}. The zero-temperature RG flow equations for the couplings 
and the self-energy can be written as\cite{Zanchi2000,Binz2003,SWTsai2005}
\begin{eqnarray}
\!\!\!\!&&\partial_{\ell}U_\ell(k_1,k_2,k_3)=\int_{p}\frac{d}{d\ell}
\big[C_\ell(p)C_{\ell}(q_{pp}-p)\big] \notag\\
\!\!\!\!&\times&\!\!\!\!U_\lppt(k_1,k_2,p)U_\lppt(k_3,k_4,p)+\int_{p}\frac{d}{d\ell}
\big[C_\ell(p)C_{\ell}(p+q'_{ph})\big] \notag\\
\!\!\!\!&\times&\!\!\!\!U_\lphpt(p,k_1,k_4)U_\lphpt(p,k_3,k_2)+\int_{p}\frac{d}{d\ell}
\big[C_\ell(p)C_{\ell}(p+q_{ph})\big] \notag\\
\!\!\!\!&\bigg[&\!-2U_\lpht(k_1,p,k_3)U_\lpht(k_4,p,k_2)
+U_\lpht(p,k_1,k_3) \notag\\
\!\!\!&\times&\!U_\lpht(k_4,p,k_2)+U_\lpht(k_1,p,k_3)U_\lpht(p,k_4,k_2)\bigg],
\label{U_l}
\end{eqnarray}
\begin{eqnarray}
\partial_\ell\Sigma_\ell(k_1)\!=\!\!\int_{p}\frac{d}{d\ell}
\left[C_\ell(p)\right]\left[U_\ell(k_1,p,p)-2U_\ell(k_1,p,k_1)\right],
\label{S_l}
\end{eqnarray}
where we have defined
$q_{pp}\equiv k_1+k_2$, $q_{ph}\equiv k_1-k_3$, $q'_{ph}\equiv k_1-k_4$, 
$\lppt\equiv min\{\ell_p,\ell_{q_{pp}-p}\}$, $\lpht\equiv
min\{\ell_p,\ell_{p+q_{ph}}\}$, and $\lphpt\equiv min\{\ell_p,\ell_{p+q'_{ph}}\}$.
We have also used the shorthand notation
$\int_p=\int_{-\infty}^{+\infty}\frac{d\omega}{2\pi}\int\frac{d^2p}{(2\pi)^2}$.
As we see, the above RG flow equations are in general non-local in RG parameter 
time $\ell$.

\begin{figure}[t]
\includegraphics[angle=-90,totalheight=5.5cm,width=7.0cm,viewport=5 5 585
685,clip]{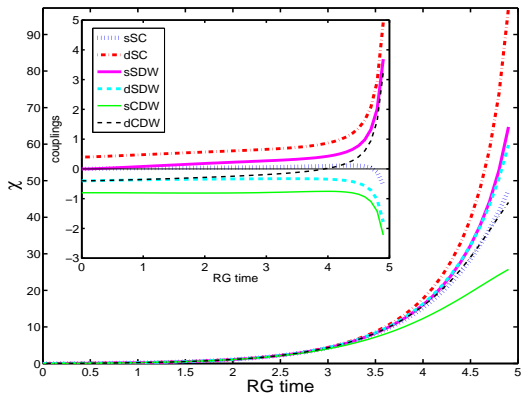} 
\caption{(Color online) RG flow of the static homogeneous susceptibilities
  for the different instabilities for $u_0=0.5$, $\lambda_0=0.6$,
  $\lambda_\pi=0.4$ and $\omega_E=1.0$. 
Inset: Corresponding RG flow of the couplings with
$\omega_1=\omega_2=\omega_3=0$. The color scheme (according to decreasing 
divergence strength of the susceptibilities) is thick red line for dSC,
thick purple line for sSDW, thick dashed cyan line for dSDW, thick doted blue
line for sSC, thin dashed black line for dCDW, and thin green line for sCDW.}
\label{Fig:RGchi}
\end{figure}
For the square lattice at half-filling the FS contains van Hove points where
the density of states is logarithmically singular. It is a well-justified 
approximation\cite{Schulz1987} to divide the FS into regions around the singular
points $\vc{Q}_1=(\pi,0)+\vc{k}$ and $\vc{Q}_2=(0,\pi)+\vc{k}$ with 
$|\vc{k}|\ll\pi$ (called two-patch approximation), where the majority of 
electronic states are expected to reside. This results in four type of e-e 
scattering processes defined as $g_1 \equiv U(\vc{Q}_1,\vc{Q}_2,\vc{Q}_2)$,
$g_2 \equiv U(\vc{Q}_1,\vc{Q}_1,\vc{Q}_1)$, $g_3 \equiv U(\vc{Q}_1,\vc{Q}_1,\vc{Q}_2)$,
$g_4 \equiv U(\vc{Q}_1,\vc{Q}_2,\vc{Q}_1)$,
which in the fully retarded case become frequency-dependent 
$g_i(\omega_1,\omega_2,\omega_3)$. Deformations of the FS have been shown to be 
stable with respect to corrections due to e-e 
interactions\cite{Gonzalez1996,FS-corrections} and need not be of a concern
in this approximation. During the RG flow the electronic density is kept fixed
which results in the chemical potential flow canceling the flow of the real part
of the self-energy. Non-locality in the RG equations is lifted because momentum
transfers can only be zero or $(\pi,\pi)$ and since
$\xi_\vc{k}=-\xi_{\vc{k}+(\pi,\pi)}$ all RG parameters map back to $\ell$.
Phononic anisotropy is introduced by distinguishing e-ph scattering processes that
involve electrons from the same patch ($g_2$, $g_4$) and those that scatter an
electron from one patch to another ($g_1$, $g_3$), and by assigning different 
coupling strengths ($\lambda_0$, $\lambda_\pi$) to them. The only place
that phonons enter into the RG flow is through the following initial conditions
\begin{eqnarray}
g_{1,3}^{\ell=0}(\omega_1,\omega_2,\omega_3) &=& u_0 - \lambda_\pi
\frac{\omega_E^2}{\omega_E^2 + (\omega_1-\omega_3)^2},\label{BC:g13}\\
g_{2,4}^{\ell=0}(\omega_1,\omega_2,\omega_3) &=& u_0 - \lambda_0
\frac{\omega_E^2}{\omega_E^2 + (\omega_1-\omega_3)^2} . \label{BC:g24} 
\end{eqnarray}
The reader interested in the analytic details of the flow equations can consult
the Appendix where for the sake of completeness we provide the exact RG flow
equations for the couplings and the self-energy.

\subsection{RG flow of the susceptibilities}
\begin{figure}[t]
\includegraphics[angle=-90,totalheight=5.5cm,width=7.0cm,viewport=5 5 570
710,clip]{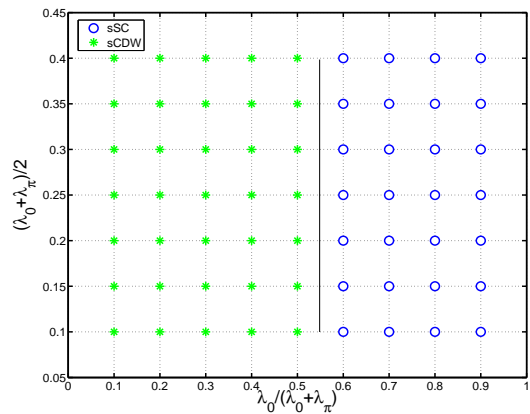} 
\caption{(Color online) 
Phase diagram ($u_0=0.0$ and $\omega_E=1.0$). Without on-site repulsion, the 
dSC phase disappears completely even for large phonon anisotropy. Color scheme
is identical to Fig.~(\ref{Fig:PhaseDiagram:1.0}).}
\label{Fig:PhaseDiagram:u0}
\end{figure}

As we mentioned in the introduction, the different instabilities associated with 
the Hubbard model at half-filling are superconductivity (sSC, dSC), 
antiferromagnetism (sSDW, dSDW), and charge density wave (sCDW, dCDW) with the 
corresponding couplings\cite{Schulz1987} (we suppress the implicit frequency 
dependence for clarity)
\begin{eqnarray}
u_{\textrm{(sd)SC}}&=&-2(g_2\pm g_3), \label{u:sdSC}\\
u_{\textrm{(sd)SDW}}&=&2(g_4 \pm g_3), \label{u:sdSDW}\\
u_{\textrm{(sd)CDW}}&=&-2(2g_1\pm g_3-g_4), \label{u:sdCDW}
\end{eqnarray}
where the signs are chosen so that strong fluctuations in a channel produce a
positive value for the corresponding coupling, irrespective of its attractive
(SC,CDW) or repulsive (SDW) nature. Since the couplings are functions of 
frequency and the most divergent ones are not necessarily at zero
frequency\cite{Tam}, only a divergent susceptibility can determine which phase
has dominant fluctuations. The corresponding order parameters for the 
different phases are 
\begin{eqnarray}
\Delta^{\textrm{SC}}(\xi,\theta,\tau) &\equiv& \sum_{\sigma} \sigma
\Psi_{\sigma,\vc{k}}(\tau)\Psi_{-\sigma,-\vc{k}}(\tau), 
\label{D:SC}\\
\Delta^{\textrm{SDW}}(\xi,\theta,\tau) &\equiv& \sum_{\sigma}
\bar{\Psi}_{\sigma,\vc{k}}(\tau)\Psi_{-\sigma,\vc{k}+(\pi,\pi)}(\tau), 
\label{D:SDW}\\
\Delta^{\textrm{CDW}}(\xi,\theta,\tau) &\equiv& \sum_{\sigma}
\bar{\Psi}_{\sigma,\vc{k}}(\tau)\Psi_{\sigma,\vc{k}+(\pi,\pi)}(\tau), 
\label{D:CDW}
\end{eqnarray}
and involve the creation of particle-particle (p-p) and particle-hole (p-h) pairs 
at given angle $\theta$ and energy $\xi$ set by $\vc{k}$. The homogeneous 
frequency-dependent susceptibilities associated with these order parameters are 
calculated by extending the one-loop RG scheme of Zanchi and 
Schulz\cite{Zanchi2000}. Their definition involves scattering processes between 
pairs at different angles $\theta$ with energies corresponding to fast modes and
are given by
\begin{eqnarray}
&\chi^{\delta}(\theta_1,\theta_2,|\tau_1-\tau_2|)=\prod_{i=1}^{2}\int d\xi_i
\Theta(|\xi_i|-\Ll)J(\xi_i,\theta_i) \notag\\
&\times\langle\Delta^{\delta}(\xi_1,\theta_1,\tau_1)\bar{\Delta}^{\delta}
(\xi_2,\theta_2,\tau_2)\rangle,
\label{chi:t}
\end{eqnarray}
where $J(\xi,\theta)$ is the Jacobian for the $\vc{k}\rightarrow(\xi,\theta)$
transformation, and $\delta=$(SC,SDW,CDW). In the two-patch approximation 
$\theta$ can take only two values and each susceptibility becomes a $2\times2$ 
matrix that we diagonalize to extract the symmetric (s-wave) and antisymmetric 
(d-wave) eigenvectors of each corresponding order. The interested reader is urged 
again to refer to the Appendix where we provide the full expressions for the 
susceptibility flow equations of all the different orders calculated to one-loop 
accuracy.

\section{Results}
\begin{figure}[t]
\includegraphics[angle=-90,totalheight=5.5cm,width=7.0cm,viewport=5 5 570
710,clip]{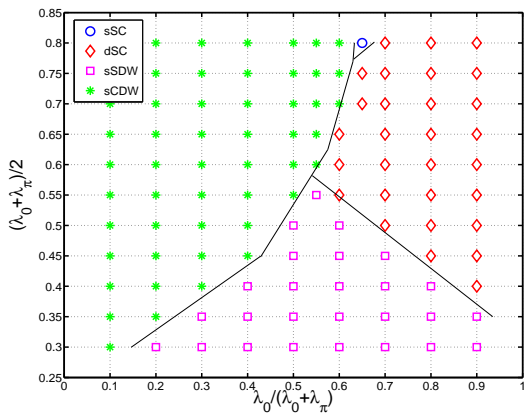} 
\caption{(Color online) Phase diagram for the same on-site repulsion
  ($u_0=0.5$) as in Fig.~(\ref{Fig:PhaseDiagram:1.0}), but for a smaller
  phonon frequency ($\omega_E=0.1$). 
  Slower phonons suppress superconductivity over the nesting-associated CDW and 
  SDW phases. Color scheme is identical to Fig.~(\ref{Fig:PhaseDiagram:1.0}).}
\label{Fig:PhaseDiagram:0.1}
\end{figure}

In Fig.~(\ref{Fig:RGchi}) we show the numerical solution of the RG flow
equations for the couplings with $\omega_1=\omega_2=\omega_3=0$
(inset) and the corresponding static ($\omega=0$) homogeneous susceptibilities 
for the case of $\lambda_0=0.6$, $\lambda_\pi=0.4$, $\omega_E=1.0$, and 
$u_0=0.5$. All parameters are expressed in energy units of $2t$ (a quarter 
of the band width) and for the numerical implementation the frequencies are 
discretized into a total of $N=41$ divisions up to a maximum value of 
$\omega_{\rm max} = 6.0$. 
The particular choice of $\lambda_0>\lambda_\pi$ enhances the attractive BCS
type pairing processes associated with $g_2$, while suppressing the dominant
repulsive nesting channels of $g_3$ (Eqs.~(\ref{BC:g13}-\ref{BC:g24})), and 
tilts the balance between the usually dominant sSDW and subdominant dSC phase.
When reversed ($\lambda_\pi>\lambda_0$), the attractive channels of $g_3$ 
combined with $g_1$ lead to a CDW instability.

The general phase diagram for fixed values of $u_0=0.5$ and $\omega_E=1.0$
is shown in Fig.~(\ref{Fig:PhaseDiagram:1.0}), where the e-ph coupling is 
parametrized by the mean value $\bar{\lambda}=(\lambda_0+\lambda_\pi)/2$ and 
relative anisotropy $r=\lambda_0/(\lambda_0+\lambda_\pi)$. Previous work on 
the phononic effects in this type of system was limited along $r=0.5$ (isotropic
phonons) and $\bar{\lambda}\gg u_0$ and found the nesting-related dominant CDW 
competing with sSC, while dSC was suppressed\cite{Berger1995}. In our generic 
study, extending to all possible configurations of coupled e-ph systems, 
we find that close to $\bar\lambda=u_0$ and along $r=0.5$ there are four 
competing phases including dSC. Deep in the repulsive region antiferromagnetism
prevails as expected (the counterpart of sCDW), but for $r>0.5$ type of 
anisotropy ($\lambda_0>\lambda_\pi$), there is a large region of dSC dominance 
(the case of Fig.~(\ref{Fig:RGchi}) is a point in that region). 
Another candidate for this parameter space is dCDW (charge flux 
phase)\cite{dcdw} but we find that while this channel 
does get renormalized significantly (Fig.~\ref{Fig:RGchi}), the suppressed 
$g_1$ and $g_3$ couplings undermine its strength.

It should be pointed out explicitly that the generic characteristics of the 
phase diagram in Fig.~(\ref{Fig:PhaseDiagram:1.0}) are independent of the actual
value of the repulsive $u_0$. The same phase diagram is always expected near 
$\bar\lambda=u_0$ and details such as adding a next-nearest neighbor hoping 
term in the Hamiltonian or doping away from half-filling will only 
$\it{enhance}$ our findings associated with dSC since all nesting-related 
processes will then be additionally suppressed. In the absence of Hubbard 
on-site repulsion ($u_0=0$), not only sSDW does not occur, as expected, but 
the dSC phase also disappears completely as shown in 
Fig.~(\ref{Fig:PhaseDiagram:u0}) for the $\omega_E=1.0$ case. Therefore, the
dSC phase is not being driven solely by anisotropic phonons, but by the combined 
effect of anisotropic phonons, on-site repulsion and nesting of the FS.
\begin{figure}[t]
\includegraphics[angle=-90,totalheight=5.5cm,width=7.0cm,viewport=65 75 780 910,clip]
{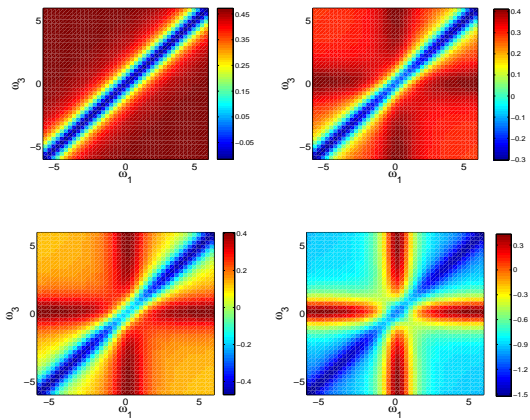}
\caption{(Color online) Evolution of $g_2(\omega_1,-\omega_1,\omega_3)$ 
coupling for $\lambda_0=0.6$, $\lambda_\pi=0.4$, and $\omega_E=1.0$.
The RG steps chosen are $\ell=0.4,2.4,3.4,4.9$ (top left, top right, 
bottom left, bottom right). The color scheme is from lower dark blue (attractive)
to higher dark red (repulsive) values.}
\label{Fig:g2m-snapshot}
\end{figure}

In order to demonstrate the importance of retardation, we show in 
Fig.~(\ref{Fig:PhaseDiagram:0.1}) the corresponding phase diagram for slower
phonons of $\omega_E=0.1$, keeping $u_0=0.5$ fixed. We see that as compared to 
Fig.~(\ref{Fig:PhaseDiagram:1.0}), the SC regions are suppressed and pushed 
towards larger values of $\bar{\lambda}$. This is a clear indication
that the internal dynamics of the coupling functions extend beyond renormalizing
the effective e-e interaction to attractive values. There is a rich internal
frequency structure among the couplings as we show in 
Fig.~(\ref{Fig:g2m-snapshot}), where we plot the evolution of the pairing
channel of $g_2(\omega_1,-\omega_1,\omega_3)$ for the case referring to the dSC 
instability of Fig.~(\ref{Fig:RGchi}). At the beginning of the flow we see
the attractive part of the coupling (limited around $\omega_3=\omega_1$), to be
much weaker compared to the repulsive parts. As the flow progresses,
and we go beyond the initial conditions of Eqs.~(\ref{BC:g13}-\ref{BC:g24}),
the repulsive parts of the coupling are suppressed in $\omega$-space and in 
actual values, while the attractive parts proliferate in both. Remarkably 
enough, when the critical point is reached we see that the same type of 
scattering process can become attractive or repulsive, depending on the dynamics 
of the electrons being scattered, with a range of strength that extends as much
as three times the bare $u_0$. This is a strong indication that retardation 
effects play a crucial and dominant role in these types of systems.

\section{Conclusions}

In conclusion, we have presented a one-loop functional renormalization group
analysis of the Hubbard model at half-filling, self-consistently including the 
electron-phonon interaction. We find that different values of the e-ph coupling
energy and anisotropy can tune the system into different instabilities,
including sSDW, sCDW, sSC, and dSC. When only the instantaneous on-site
repulsion is present, sSDW is well known to dominate. In the absence of
on-site repulsion, phonon mediated attraction drives the system into 
sCDW and sSC phases. In the presence of both interactions, depending on the 
competition between them, these phases appear in different parts of the phase 
diagram. In addition, these interactions also {\it cooperate} to generate a dSC 
phase. This phase therefore only emerges because of the interplay between the 
physics of Coulomb interactions and phonons. Retardation effects play an 
important role in the onset of these phases and also determine the size of the 
different regions in the phase diagram.

Our work is not to be considered as a complete study of the two-dimensional 
Hubbard model in the presence of the e-ph interaction but as a preliminary but
self-consistent study that hints towards the right physics of such a system. 
In other words, what the functional RG study can beautifully provide to us is an 
unbiased study of the instabilities associated with this type of system in a 
self-consistent manner to the given one-loop accuracy order, which allows us the 
flexibility of probing an immense parameter phase space in a rather inexpensive 
way (numerically and analytically). The assumptions of this approach are the 
weak coupling regime in the electron-phonon interaction and that the largest 
energy scale of the system is the bandwidth. Within this approximate method we 
are able to draw the first charts of the rich phase space when the fermion-boson 
interaction is no longer neglected but self-consistently included as well.

\acknowledgments We thank A. H. Castro Neto, D. K. Campbell, J. B. Marston,
R. Shankar, and K.-M. Tam for useful discussions.

\appendix

\section{Coupling and self-energy flow equations}

The renormalization-group flow equations for the four couplings and the 
self-energy are directly derived from Eqs.~(\ref{U_l}-\ref{S_l}) when
the two-patch approximation is used which maps all momenta transfers
either to $(0,0)$ or $(\pi,\pi)$ converting the intrinsically non-local
flow equations to local. This procedure is very well presented in the work of
Zanchi and Schulz\cite{Zanchi2000} for the reader interested in the full 
exposition of details. Here we only highlight the basic points along with 
presenting the final formulas for the sake of completeness.

In the two-patch approximation the whole FS is divided into two patches 
(each patch has its redundant mirror image). The RG flow 
Eqs.~(\ref{U_l}-\ref{S_l}) involve a delta function constraining the electronic
energies $\ksik$ to be ``on-shell" which can be above $(+\Ll)$ or below $(-\Ll)$
the FS. This constraint induces a one-to-one correspondence between $\vc{k}$
and $(\theta,|\ksik|=\Ll)$ which consequently is employed to simplify the 
2D-$\vc{k}$ integration into an azimuthal integration over $\theta$, with
the proper Jacobian $J(\theta,\ksik)$ introduced. The integrand
is $\theta$-independent and one can define a general operator $\hat{F}_{\nu}$ 
acting on any product of two coupling functions  
$g_i(\omega_1,\omega,\omega_3)g_j(\omega_2,\omega,\omega_4)$ according to
\begin{widetext}
\begin{eqnarray}
&\hat{F}_{\nu}&\!\![\Omega][g_ig_j](\omega_1,\omega,\omega_3)(\omega_2,\omega,\omega_4) 
\equiv\frac{\Ll}{\pi^3}\int_{0}^{\pi/4}d\theta\frac{k(\theta,-\Ll)}
{\big|\frac{d\xi}{dk}(\theta,-\Ll)\big|}\int_{-\infty}^{+\infty}d\omega 
g_i(\omega_1,\omega,\omega_3)g_j(\omega_2,\omega,\omega_4) \notag\\
&\times&\frac{\big(\omega-\Sigma_\ell''(\omega)\big)\big(\Omega-
\Sigma_\ell''(\Omega)\big)+\nu\Ll^2}{\big[\big(\omega-\Sigma_\ell''(\omega)\big)
\big(\Omega-\Sigma_\ell''(\Omega)\big)+\nu\Ll^2\big]^2+\Ll^2\big[\omega
-\Sigma_\ell''(\omega)-\nu\big(\Omega-\Sigma_\ell''(\Omega)\big)\big]^2},
\label{FnuDefinition}
\end{eqnarray}
where $\Omega$ depends on the external frequencies and $\omega$ (which is 
integrated out), $\nu=\pm$ and effectively defines two types of $\hat{F}$ 
operators, while $\Sigma_\ell''$ is the imaginary part of the self-energy. The
Jacobian at half-filling has the convenient property 
$\oint J(\theta,-\Ll)=\oint J(\theta,\Ll)$. The complete RG flow equations for 
the four couplings and the imaginary part of the self-energy can then be written 
as
\begin{eqnarray}
&&\!\!\!\!\!\frac{\partial g_1}{\partial\ell}(\omega_1,\omega_2,\omega_3)=
\hat{F}_+[\omega+\omega_1-\omega_3]
\bigg(
-2[g_3g_3+g_1g_1](\omega_1,\omega,\omega_3)(\omega_4,\omega,\omega_2)
+[g_3g_3+g_1g_4](\omega_1,\omega,\omega_3)(\omega,\omega_4,\omega_2) \notag\\
&+&[g_3g_3+g_4g_1](\omega,\omega_1,\omega_3)(\omega_4,\omega,\omega_2)
\bigg)+\hat{F}_+[\omega+\omega_3-\omega_1]
\bigg(
-2[g_3g_3+g_1g_1](\omega_3,\omega,\omega_1)(\omega_2,\omega,\omega_4) \notag\\
&+&[g_3g_3+g_1g_4](\omega_3,\omega,\omega_1)(\omega,\omega_2,\omega_4)
+[g_3g_3+g_4g_1](\omega,\omega_3,\omega_1)(\omega_2,\omega,\omega_4)
\bigg),
\label{g1flow}
\end{eqnarray}
\begin{eqnarray}
&&\!\!\!\!\!\frac{\partial g_2}{\partial\ell}(\omega_1,\omega_2,\omega_3)=
\hat{F}_-[\omega_1+\omega_2-\omega]
\bigg(
[g_3g_3+g_2g_2](\omega_1,\omega_2,\omega)(\omega_3,\omega_4,\omega)
+[g_3g_3+g_2g_2](\omega_2,\omega_1,\omega)(\omega_4,\omega_3,\omega)
\bigg)
\label{g2flow}
\end{eqnarray}
\begin{eqnarray}
&&\!\!\!\!\!\frac{\partial g_3}{\partial\ell}(\omega_1,\omega_2,\omega_3)=
\hat{F}_+[\omega+\omega_1-\omega_3]
\bigg(
-2[g_3g_1+g_1g_3](\omega_1,\omega,\omega_3)(\omega_4,\omega,\omega_2)
+[g_3g_4+g_1g_3](\omega_1,\omega,\omega_3)(\omega,\omega_4,\omega_2) \notag\\
&+&[g_3g_1+g_4g_3](\omega,\omega_1,\omega_3)(\omega_4,\omega,\omega_2)
\bigg)+\hat{F}_+[\omega+\omega_3-\omega_1]
\bigg(
-2[g_3g_1+g_1g_3](\omega_3,\omega,\omega_1)(\omega_2,\omega,\omega_4) \notag\\
&+&[g_3g_4+g_1g_3](\omega_3,\omega,\omega_1)(\omega,\omega_2,\omega_4)
+[g_3g_1+g_4g_3](\omega,\omega_3,\omega_1)(\omega_2,\omega,\omega_4)
\bigg) \notag\\
&+&\hat{F}_-[\omega_1+\omega_2-\omega]
\bigg(
[g_2g_3+g_3g_2](\omega_1,\omega_2,\omega)(\omega_3,\omega_4,\omega)
+[g_2g_3+g_3g_2](\omega_2,\omega_1,\omega)(\omega_4,\omega_3,\omega) 
\bigg) \notag\\
&+&\hat{F}_+[\omega+\omega_2-\omega_3]
[g_3g_4+g_4g_3](\omega,\omega_2,\omega_3)(\omega,\omega_4,\omega_1)
+\hat{F}_+[\omega+\omega_3-\omega_2]
 [g_3g_4+g_4g_3](\omega,\omega_3,\omega_2)(\omega,\omega_1,\omega_4),
\label{g3flow}
\end{eqnarray}
\begin{eqnarray}
&&\!\!\!\!\!\frac{\partial g_4}{\partial\ell}(\omega_1,\omega_2,\omega_3)=
\hat{F}_+[\omega+\omega_2-\omega_3]
[g_3g_3+g_4g_4](\omega,\omega_2,\omega_3)(\omega,\omega_4,\omega_1) \notag\\
&+&\hat{F}_+[\omega+\omega_3-\omega_2]
[g_3g_3+g_4g_4](\omega,\omega_3,\omega_2)(\omega,\omega_1,\omega_4),
\label{g4flow}
\end{eqnarray}
\begin{eqnarray}
%\frac{\partial\Sigma_\ell'}{\partial\ell}(\omega_1)&=&
%\frac{\Ll^2}{\pi^3}\int_{0}^{\pi/4}d\theta\frac{k(\theta,-\Ll)}{\big|\frac{d\xi}
%{dk}(\theta,-\Ll)\big|}\int_{-\infty}^{+\infty}d\omega\frac{[g_1+g_2]
%(\omega_1,\omega,\omega)-2[g_2+g_4](\omega_1,\omega,\omega_1)}
%{\big(\omega-\Sigma_\ell''(\omega)\big)^2+\Ll^2}.
%\label{Spflow}\\
\frac{\partial\Sigma_\ell''}{\partial\ell}(\omega_1)&=&
\frac{\Ll}{\pi^3}\int_{0}^{\pi/4}d\theta\frac{k(\theta,-\Ll)}{\big|\frac{d\xi}
{dk}(\theta,-\Ll)\big|}\int_{-\infty}^{+\infty}d\omega\frac{\big([g_1+g_2]
(\omega_1,\omega,\omega)-2[g_2+g_4](\omega_1,\omega,\omega_1)\big)
\big(\omega-\Sigma_\ell''(\omega)\big)}{\big(\omega-\Sigma_\ell''(\omega)
\big)^2+\Ll^2}.
\label{Sppflow}
\end{eqnarray}
\end{widetext}
The above equations are numerically solved for each RG step $\ell$ until any
coupling for any frequency channel diverges to values greater than
$20t$ at which point we stop the algorithm and form all the couplings
associated with the different instabilities given by 
Eqs~(\ref{u:sdSC}-\ref{u:sdCDW}). For frequency independent interactions the 
above equations reduce to the usual\cite{Schulz1987}
\begin{eqnarray}
\frac{\partial g_1}{\partial\ell}&=&-2g_1(g_1-g_4),\\
\frac{\partial g_2}{\partial\ell}&=&-g_2^2-g_3^2,\\
\frac{\partial g_3}{\partial\ell}&=&-2g_3(g_1+g_2-2g_4),\\
\frac{\partial g_4}{\partial\ell}&=&g_3^2+g_4^2.
\end{eqnarray}

\section{Susceptibility flow equations}

Once the RG flow for the couplings is numerically solved and the divergence 
point towards strong coupling is reached we calculate the susceptibilities
associated with the major orders in the system. As we mentioned in the text,
a general susceptibility calculation involves Eq.~(\ref{chi:t}) for all different 
orders. By including (to one-loop) all RG vertex corrections\cite{Zanchi2000} we 
obtain the general homogeneous susceptibility flow equation which in the two-patch
approximation reduces to a $2\times 2$ matrix given by
\begin{eqnarray}
&&\!\!\!\!\!\!\partial_\ell\chi^{\delta}_\ell\!(\omega_1)\!=\!
\frac{\Ll}{\pi^3}\!\!\int_{0}^{\pi/4}\!\!\!\!d\theta J(\theta,-\Ll)\!\!
\int_{-\infty}^{+\infty}\!\!\!\!d\omega W_\ell^{\delta}(\omega_1,\omega)\notag\\
&\times&\begin{pmatrix} 
z_{11}z_{11}+z_{12}z_{21} & z_{11}z_{12}+z_{12}z_{22} \\
z_{21}z_{11}+z_{22}z_{21} & z_{22}z_{22}+z_{21}z_{12}
\end{pmatrix},
\label{chi:delta}
\end{eqnarray}
where $\delta=(AF,SC,CDW)$ and we have used the shorthand notation 
$z_{ij}z_{lm}=z^{\delta}(\theta_i,\theta_j;\omega_1,\omega)
z^{\delta}(\theta_l,\theta_m;\omega,\omega_1)$ and the definitions
\begin{eqnarray}
&&\!\!\!\!\!\!\!\!W_\ell^{AF}(\omega_1,\omega)=W_\ell^{CDW}(\omega_1,\omega)= \notag\\
&&\!\!\!\!\!\!\!\!\frac{\Ll^2\!+\!\big(\omega\!-\!\Sigma''_\ell(\omega)\big)
\big(\omega\!+\!\omega_1\!-\!\Sigma''_\ell(\omega\!+\!\omega_1)\big)}
{\big[\big(\omega\!-\!\Sigma''(\omega)\big)^2\!\!\!+\!\!\Ll^2\big]
\big[\big(\omega\!+\!\omega_1\!-\!\Sigma''_\ell(\omega\!+\!\omega_1)\big)^2
\!\!\!+\!\!\Ll^2\big]},
\label{W:AF-CDW}
\end{eqnarray}
\begin{eqnarray}
&&\!\!\!\!\!\!\!\!W_\ell^{SC}(\omega_1,\omega)= \notag\\
&&\!\!\!\!\!\!\!\!\frac{\Ll^2\!-\!\big(\omega\!-\!\Sigma''_\ell(\omega)\big)
\big(\omega_1\!-\!\omega\!-\!\Sigma''_\ell(\omega_1\!-\!\omega)\big)}
{\big[\big(\omega\!-\!\Sigma''(\omega)\big)^2\!\!\!+\!\!\Ll^2\big]
\big[\big(\omega_1\!-\!\omega\!-\!\Sigma''_\ell(\omega_1\!\!-\!\omega)\big)^2
\!\!\!+\!\!\Ll^2\big]}.
\label{W:SC}
\end{eqnarray}
The RG flow equations for the vertex functions associated with the nested 
related phases are given by
\begin{eqnarray}
&&\!\!\!\!\!\partial_\ell z^{AF}_\ell\!\!(\omega_1,\omega_2)\!=\!
\frac{\Ll}{\pi^3}\!\!\int_{0}^{\pi/4}\!\!\!\!d\theta J(\theta,-\Ll)\!\!
\int_{-\infty}^{+\infty}\!\!\!\!d\omega W_\ell^{AF}(\omega_1,\omega) \notag\\
&&\times\begin{pmatrix} 
z_{11}g_3+z_{12}g_4 & z_{11}g_4+z_{12}g_3 \\
z_{21}g_3+z_{22}g_4 & z_{21}g_4+z_{22}g_3
\end{pmatrix},
\label{z:AF}
\end{eqnarray}
and 
\begin{eqnarray}
&&\!\!\!\!\!\!\!\partial_\ell z^{CDW}_\ell\!\!(\omega_1,\omega_2)\!\!=\!\!
\frac{-\Ll}{\pi^3}\!\!\!\int_{0}^{\pi/4}\!\!\!\!\!\!\!\!d\theta J(\theta,-\Ll)
\!\!\!\int_{-\infty}^{+\infty}\!\!\!\!\!\!\!\!d\omega W_\ell^{CDW}\!\!
(\omega_1,\omega) \notag\\
&&\times\begin{pmatrix} 
z_{11}g_\alpha+z_{12}g_\beta & z_{11}g_\beta+z_{12}g_\alpha \\
z_{21}g_\alpha+z_{22}g_\beta & z_{21}g_\beta+z_{22}g_\alpha
\end{pmatrix},
\label{z:CDW}
\end{eqnarray}
where $z_{ij}g_k\equiv z^{\delta}(\theta_i,\theta_j;\omega_1,\omega)
g_k(\omega_2,\omega_1+\omega,\omega)$ and we used the additional definitions
\begin{eqnarray}
\!\!\!\!\!\!\!\!\!g_\alpha\!(\omega_2,\omega_1\!\!+\!\!\omega,\omega)
\!\!\!&\equiv&\!\!\!2g_3(\omega_1\!\!+\!\!\omega,\omega_2,\omega)\!\!
-\!\!g_3(\omega_2,\omega_1\!\!+\!\!\omega,\omega), \\
\!\!\!\!\!\!\!\!\!g_\beta\!(\omega_2,\omega_1\!\!+\!\!\omega,\omega)
\!\!\!&\equiv&\!\!\!2g_1(\omega_1\!\!+\!\!\omega,\omega_2,\omega)\!\!
-\!\!g_4(\omega_2,\omega_1\!\!+\!\!\omega,\omega).
\end{eqnarray}
For the SC-related vertex function we have
\begin{eqnarray}
&&\!\!\!\!\!\partial_\ell z^{SC}_\ell\!(\omega_1,\omega_2)\!=\!
\frac{\Ll}{\pi^3}\!\!\int_{0}^{\pi/4}\!\!\!\!\!d\theta J(\theta,-\Ll)\!\!
\int_{-\infty}^{+\infty}\!\!\!\!\!d\omega W_\ell^{SC}(\omega_1,\omega) \notag\\
&&\times\begin{pmatrix} 
z_{11}g_2+z_{12}g_3 & z_{11}g_3+z_{12}g_2 \\
z_{21}g_2+z_{22}g_3 & z_{21}g_3+z_{22}g_2
\end{pmatrix}.
\label{z:SC}
\end{eqnarray}
where $z_{ij}g_k\equiv z^{SC}(\theta_i,\theta_j;\omega_1,\omega)
g_k(\omega,\omega_1-\omega,\omega_2)$. All susceptibilities are zero at the 
initial RG step $\ell=0$, and the vertex functions obey 
\begin{equation}
z^{\delta}_{\ell=0}(\theta_1,\theta_2;\omega_1,\omega_2)=\frac{1}{4}
\delta_{\theta_1,\theta_2}\delta(\omega_1-\omega_2).
\end{equation}
Also, due to the fact that we are at half-filling the flow equations for the 
nested related phases are local\cite{Zanchi2000}.

\end{document}